 \documentclass[preprint]{aastex}

\shorttitle{Kepler LPVs}
\shortauthors{Hartig et al.}

\begin{document} 

\title{Kepler and the Long Period Variables}


\author{Erich Hartig\footnotemark  \footnotetext{Joint lead author}}
\affil{University of Vienna, Department of Astrophysics, 
T\"urkenschanzstrasse 17, A-1180 Vienna, Austria}
\email{erich.hartig@univie.ac.at}

\author{Jennifer Cash$^1$}
\affil{Department of Biological and Physical Sciences, 
South Carolina State University, 
P.O. Box 7024,
Orangeburg, SC 29117,USA}
\email{jcash@physics.scsu.edu} 

\author{Kenneth H. Hinkle}
\affil{National Optical Astronomy Observatories\\
P.O. Box 26732, Tucson, AZ 85726 USA}
\email{hinkle@noao.edu}

\author{Thomas Lebzelter}
\affil{University of Vienna, Department of Astrophysics, 
T\"urkenschanzstrasse 17, A-1180 Vienna, Austria}
\email{thomas.lebzelter@univie.ac.at} 

\author{Kenneth J. Mighell}
\affil{National Optical Astronomy Observatories\\
P.O. Box 26732, Tucson, AZ 85726 USA}
\email{mighell@noao.edu}

\and

\author{Donald K. Walter}
\affil{Department of Biological and Physical Sciences, 
South Carolina State University, 
P.O. Box 7296,
Orangeburg, SC 29117,USA}
\email{dkw@physics.scsu.edu}

\begin{abstract}
High precision Kepler photometry is used to explore the details of
AGB light curves.  Since AGB variability has a typical time scale
on order of a year we discuss at length the removal of long term
trends and quarterly changes in Kepler data.  Photometry for a small
sample of nine SR AGB stars are examined using a 30 minute cadence
over a period of 45 months.  While undergoing long period variations
of many magnitudes, the light curves are shown to be smooth at the
millimagnitude level over much shorter time intervals.  No flares
or other rapid events were detected on the sub-day time scale.  The
shortest AGB period detected is on the order of 100 days.  All 
the SR variables in our sample are shown to have multiple 
modes.  This is always the first overtone typically combined with
the fundamental.  A second common characteristic of SR variables
is shown to be the simultaneous excitation of multiple closely
separated periods for the same overtone mode.  Approximately half
the sample had a much longer variation in the light curve, likely a
long secondary period.  The light curves were all well represented
by a combination of sinusoids.  However, the properties of the
sinusoids are time variable with irregular variations present at
low level.  No non-radial pulsations were detected.  It is argued
that the long secondary period variation seen in many SR variables
is intrinsic to the star and linked to multiple mode pulsation.
\end{abstract}




\section{INTRODUCTION}

The General Catalog of Variable Stars lists multiple groups of
pulsating variable stars of late spectral type with periods of several tens
of days or longer.  While historical divisions by period and spectral
type were made in the absence of knowledge about stellar evolution,
we now recognize that these long period variable (LPV) stars are
evolved low- and intermediate-mass stars. During the red giant
branch (RGB), asymptotic giant branch (AGB), and post-AGB phases
stars are characterized by highly extended stellar envelopes providing
the basis for long-period pulsation.

The variability time scale of LPVs requires monitoring over years
to decades to properly describe the light changes.  In the 1990s
long term, rapid cadence monitoring of variable stars in selected
fields occurred as a byproduct of ground-based photometric surveys
searching for microlensing events.  This data was transformative,
providing large data sets of high quality light curves for stars
in the Magellanic Clouds \citep{wood_et_al_1999, lebzelter_et_al_2002,
kiss_bedding_2003, ita_et_al_2004}.  The Magellanic Cloud variables
are at a known distance, and hence have known luminosities, permitting
the construction of period-luminosity (P-L) diagrams for a large
sample of late-type stars.  While previously a P-L relation had
been known for the large amplitude LPVs, the Miras
\citep{glass_lloyd_evans_1981}, period-luminosity relations for all
late-type variables are apparent in the LMC data
\citep{soszynski_et_al_2009}.

In this paper we focus on those LPVs that are luminous with periods
of $\sim$100 days or longer but are not Miras.  In the parlance of
the \citet{wood_et_al_1999} P-L relation these are sequence C and
C' \citep{soszynski_wood_2013}.  In the classical variable star
nomenclature these are semi-regular (SR) variables.  The GCVS
classification is unfortunately blurred across a large range of
temperature and luminosity with SR variables found across a large
part of the cool star P-L diagram.  Here we will discuss those that
have AGB luminosity.  Other recent papers have discussed SR variables
that are less luminous and are on the RGB
\citep{christensen-dalsgaard_et_al_2001, stello_et_al_2014}.

The improved P-L diagram results in new astrophysical challenges.
The C and C' variables are pulsating radially in the fundamental
mode, a low overtone mode, or a combination of modes.  The origin
of AGB semi-regular behavior, both in cycle length and amplitude,
remains unclear, but their light curves can in some cases be
reproduced by the combination of several periods
\citep[e.g.][]{kiss_et_al_1999, kerschbaum_et_al_2001}.  The major
challenge, however, is an explanation of the long secondary periods (LSP)
found in many of the overtone pulsators. With a typical time scale
of several hundreds to thousands of days the secondary periods are
significantly longer than the fundamental mode periods and therefore
can not be explained by radial pulsation. Various solutions have
been discussed, but a satisfactory explanation has yet to be found
\citep[e.g.][]{nicholls_et_al_2009}.

One route to further understanding of these stars is precision,
continuous photometric monitoring. Recent studies faced the typical
problems of ground-based observations, time series of very limited
length, of limited photometric accuracy, and observational gaps
resulting from both diurnal and seasonal cycles. A space mission
obtaining photometry continuously over a time span of several years
is, therefore, clearly relevant.  One of the many side benefits of
the Kepler mission\footnote{http://kepler.nasa.gov/} was the

ortunity to obtain long and precise time series of variables,
including LPVs with uninterrupted sampling extending over months.

\citet{Banyai13} presented the first extensive study on the variability
of both RGB and AGB M giants using the long and continuous light
curves provided by Kepler.  In that paper the authors focused on
the global characteristics of the light variations of these cool
giants. Their study confirmed the presence of several pulsation
modes based on the patterns seen in the Petersen diagram and it
revealed a clear distinction between solar like oscillations and
large amplitude, mira-like pulsations.  This transition seems to
appear at log P(days) $\sim$ 1.

The Kepler mission was optimized to search for planet transits, not
slow variations spanning weeks to years.  As a consequence, the aim
of our paper is twofold. In the first part, we investigate the
applicability of the Kepler data products to the construction of
long base line light curves by combining individual data-sets of three
months length.  We will focus on a small sample of 12 stars, three
of which have small amplitudes that are included especially to
identify instrument specific effects. The light curves are then
used in the second part of the paper to discuss the light change
of SR variables on the basis of these uninterrupted and photometrically
precise light curves.


\section{THE SAMPLE}

The definition of the SR variable class limits the visual amplitude
to less than 2.5 magnitudes. The mira variables are defined as
having visual amplitude in excess of 2.5 magnitudes.   Thus the SR
class excludes the large amplitude, tip-AGB variables.  The intent
of our survey was to study AGB SR variables over as full a range
of variability as possible.  Unfortunely, most SR variables have not
been extensively studied in the past.  The better known objects
have been assigned a subclass a or b reflecting a more or less
regular light curve.  However, many late-type stars have time series
of ground based photometry that are too short or otherwise inadequate
to classify the variability.  The SRb class has been associated
with LSP phenomenon.  A sample spanning these variability classes
was of special interest with the goal of seeing if actual differences
existed between the subclasses.

The General Catalogue of Variable Stars \citep[GCVS][]{GCVS} list
of variables in Lyrae and Cygnus was searched for late-type SR
variables in the Kepler field.  Other surveys, for instance the
ASAS catalog \citep{ASAS}, sample a much shorter time interval and
were less useful in selecting a variety of variables.  The two
brightest SRs in the Kepler field are AF Cyg, an SRb with
spectroscopically and photometrically studied LSP
\citep{hinkle_et_al_2002}, and AW Cyg, a carbon-rich SRb variable.
Observation of these two bright stars required special target pixel
masks and hence were not included in our original target list.
However, \citep{Banyai13} selected AF Cyg as a target and developed
a special pixel mask.  We included that data.  To represent the SRa
class we identified V588 Lyr, V1670 Cyg, and V1766 Cyg.  For the
SRb class V1953 Cyg, EG Lyr, and BU Lyr were selected.  We included
two SR stars without sub-classifications, V1253 Cyg and V2412 Cyg.

Three low amplitude variable stars located in the field of the old
open cluster NGC 6791 were analyzed to test the reduction process.
The stars have periods in the range $\sim$10 - 20 d \citep{MSK2003,
BGT2003, MSS2005, MPM2007} with amplitudes of less than 0.1 magnitude.
The brevity of the ground based NGC 6791 survey results in fairly
uncertain periods.  The General Catalogue of Variable Stars
\citep[GCVS][]{GCVS} classifies the test stars as red giants with
semi-regular periods, i.e.  stars that are likely members of sequence
A or B in the LPV L-P diagram.  The spectral types are correspondingly
earlier than for the program stars with the test stars probably on
the RGB. 

Spectroscopy was carried out to determine the spectral types for
the brighter of the program stars.  The observations were made at
the Kitt Peak National Observatory using the 4-meter Mayall telescope
in May 2011 and the Coude Feed telescope in October 2010, March
2011 and May 2011.  The resulting spectral type classifications can
be found in Tables \ref{t:star_list}.  When possible
these are compared to the classifications found in the General
Catalog of Variable Stars (GCVS).  V616 Lyr was observed with the
4-meter only, V588 with both the 4-meter and the Coude Feed, and
the other observations are from the Coude Feed.  Spectral types and
luminosity classes were assigned based on \citet{Jacoby84}.

The spectral types for V616 Lyr, V588 Lyr, V1766 Cyg and V1953 Cyg
are presented here for the first time.  Our work compares favorably
with the types in the GCVS for EG Lyr.  However, we classified BU
Lyr as M4-M5 III while the GCVS lists it as an M7 star.  This could
result from changes in the spectral type with the pulsation phase.
This has been reported for semiregular variables by \citep{Percy07}.
Phase dependent changes can be mitigated by repeated observations.
EG Lyr, BU Lyr, and V1766 Cyg were observed on two or more dates
separated by several weeks or months.

Six stars in Table \ref{t:star_list} were too faint
on the dates of our observations for spectroscopic work.  Spectral
types were derived from V-K colors.  The reddening is not known for
these stars so the corresponding spectral temperatures are likely
too cool.  These data show that all the program SR stars are, as
expected, M stars (Table \ref{t:star_list}).  The low amplitude
test objects (Table \ref{t:star_list}) are earlier spectral types.

\begin{deluxetable}{rclcc}
\tabletypesize{\footnotesize}
\tablecaption{Program Stars\label{t:star_list}}
\tablewidth{0pt}
\tablehead{
\colhead{Kepler} & \colhead{GCVS} & \colhead{Var.type} & \colhead{Sp.Type} & \colhead{Sp.Type} \\
\colhead{ID}    & \colhead{Name} & \colhead{(GCVS)} & \colhead{(this study)} & \colhead{(literature)}  \\
}
\startdata
2437359  & V616 Lyr\tablenotemark{b}  & SRS     & G9\,III            \\
2569737  & V607 Lyr\tablenotemark{b}  & \nodata & K0\tablenotemark{a}\\
2570059  & V621 Lyr\tablenotemark{b}  & SRS     & G8\tablenotemark{a}\\
3431126  & EG Lyr    & SRb     & M4-5\,III                & M5\,III  \\
5176879  & BU Lyr    & SRb     & M4-5\,III                & M7       \\
5614021  & V588 Lyr  & SRa     & M2\,IIIe                 & \nodata  \\
6127083  & V1670 Cyg & SRa     & M4\,III\tablenotemark{a} & \nodata  \\
7842386  & V1766 Cyg & SRa     & M3-5\,III                & \nodata  \\
8748160  & V1253 Cyg & SR:     & M6\,III\tablenotemark{a} & \nodata  \\
9528112  & AF Cyg    & SRb     & M5\,III\tablenotemark{a} & M4 III   \\
10034169 & V2412 Cyg & SR      & M7\,III\tablenotemark{a} & \nodata  \\
12215566 & V1953 Cyg & SRb     & M3\,III                  & \nodata  \\
\enddata
\tablenotetext{a}{from NOMAD V-K using the spectral
type -- infrared color calibration of \citet{tokunaga_2000}.}
\tablenotetext{b}{Red giant SR included as a data reduction test}
\end{deluxetable}


\section{KEPLER DATA AND ANALYSIS} \label{s:dataanalysis} 

The Kepler Space telescope has an effective aperture of 0.95-meter,
a primary mirror with 1.4 meter diameter, and a field-of-view (FOV)
of 105 deg$^{2}$. It was positioned in an Earth-trailing heliocentric
orbit with a 372.5 days period.  The data discussed here is from
the original Kepler mission to observe a section of the Cygnus
region along the Orion arm centered on galactic coordinates
(76.32\,$^{\circ}$,+13.5\,$^{\circ}$) or RA=19:22:40, Dec=+44:30:00
(2000).  The photometer is an array of 42 charge coupled devices
(CCDs), with each 50x25 mm CCD having 2200x1024 pixels.  To prevent
saturation the CCDs were read out every 6 seconds.  The Kepler
observatory has two observing modes, either the short cadence mode
(SC) recording one image every 0.98 minutes or the long cadence
mode (LC) recording one observation every 29.4 minutes.  The data
were downloaded in quarterly blocks divided in monthly sections.
After each of the quarterly blocks the satellite was rotated so
that the next quarter was observed with a different part of the CCD
array.  For our project we requested only the LC mode.

For the LC mode various data products can be downloaded.  Details
can be found in the Kepler Archive
Manual\footnote{http://archive.stsci.edu/kepler/manuals/archive\_manual.pdf}.
Here is a brief summary of the data typically available for a star
observed with Kepler:

\begin{itemize}

\item Simple Aperture Photometry (SAP): In the Kepler Asteroseismic 
Science Consortium (KASC) archive these
data are called RAW.  The SAP data come with no correction of
differences in the flux zero point between the quarterly observation
blocks. Such differences stem from changes in the CCD pixels used, the use
of different target masks, changes of the seasonal thermal conditions 
causing focus changes and changes of the zodiacal light.

\item Pre-search Data Conditioning (PDC or PDCSAP): These are
preliminary light curves provided by the Kepler Science Operations
Center Science Processing Pipeline. For searching for transiting
planets various errors including systematic errors are corrected,
flux outliers removed, and data gaps partially filled.  The pipeline
producing the PDC data removes all long time trends, so that these
data are of limited use for the analysis of LPVs.

\item Target Pixel Data (TPD):  The TPD 
deliver flux information for each pixel in the window on the
CCD used to obtain the star's brightness. The size of the window
is neither the same for all stars nor necessarily constant for a
given object. Formats between 4$\times$3 to 13$\times$59 pixels have been
found for our sample stars. For the bright target AF Cyg a maximum
window size of 15$\times$587 pixels was used.

\item Superstamps: For two areas in the Kepler field, around the open clusters
NGC 6791 and NGC 6819, a complete image is provided in LC
mode. These ``superstamps'' are provided with the same 
sampling frequency as the TPD data.

\item Full Field Images (FFI): These were taken monthly at the
end of an observation cycle.

\end{itemize}

Corrected flux light curves by the KASC working groups using
their own specific pipelines for the data reduction are sporadically
available.


\subsection{Removing instrumental effects}

\subsubsection{Cotrending}

To examine the shorter period astrophysical signals the instrumental
effects that dominate the light curves first must be removed.  In
addition to zero point differences (above) there are also trends
in the flux values within a quarter.  One technique for removing
these effects is  `cotrending', described in detail by
\citet{smith_et_al_2012}, and based on the identification of common
trends in a selected set of light curves on a given chip.  Cotrending
is most effective at removing instrumental signal while preserving
astrophysical signal for periods of a few tens of days.  For the
LPVs in our set, with typical periods generally greater than 100
days, limitations of the cotrending algorithm preclude reliable
results.  The alternative way is to leave the trends in the light
curve and try to identify and remove them during period search.

Tests with low amplitude variables show that the first three
cotrending basis vectors for each individual quarter result in a
good match to the instrumentally dominated overall shape of the
light curve and the thermal features.  This gives us confidence
that cotrending is removing instrumental effects but not introducing
significant errors into the light curve.  Introduction of the fourth
and fifth basis vector start to show overfitting of the light curves
and loss of astronomical signal.  However, once the individual
quarters have been cotrended a significant difference between the
median flux levels for each quarter remains. A significant limitation
of cotrending for the LPV Kepler light curves lies in preserving
and identifying very long period variations where the method typically
results in ambiguous conclusions.

\subsubsection{Linking the quarters}\label{quarters}

In contrast to other variable classes, the study of long period variables
with Kepler requires a very precise linking of the various quarters.
Errors introduced in linking the quarters can have a major impact
on the derived long periods and their amplitudes.  Unfortunately,
this linking is hampered by the gaps of 0.9 to 3~days in time
coverage that occur between the quarters. In some cases, where the
star has not been observed continuously in all quarters, very large
gaps can occur.  Therefore, we focused in our study on the various
approaches to close the gap between the quarters correctly and, if
possible, in an objective, unambiguous, and easily reproducible
way.

First, we consider SAP data. A first simple approach is to simply
stick together the various quarters starting with the first one,
always using the flux of the last data point of the previous quarter
and the first data point of the following quarter. If cotrending
was applied first, scaling of each quarter to the same median level
can be used to reduce the discontinuities between the quarters.
This approach is particularly helpful for faint targets where the
increased scatter in the light curve complicates the scaling.

\begin{figure}[htb!]
\centering%
\includegraphics[width=0.8\textwidth]{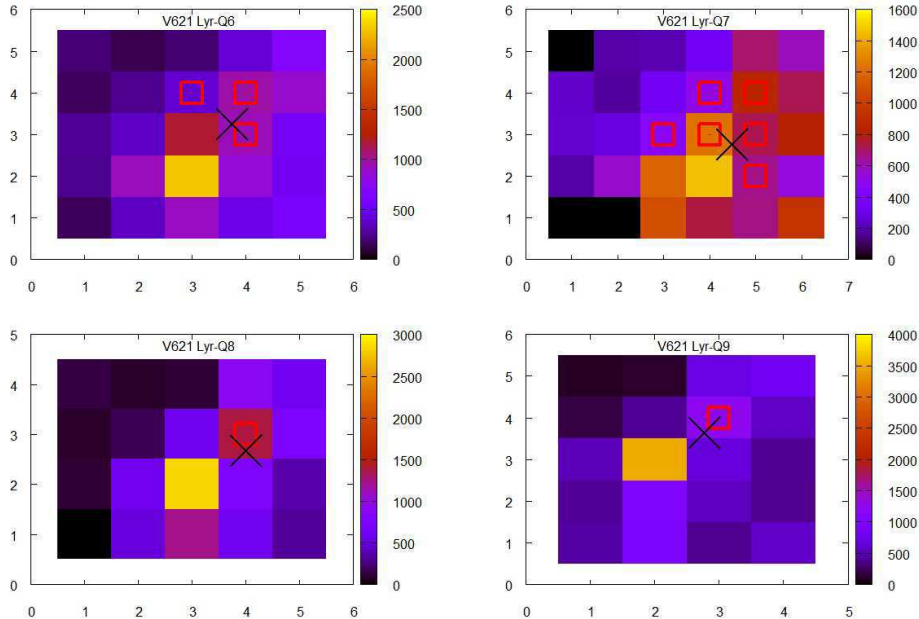}
\caption{V621 Lyr, KIC 2570059 Pixel Map of Q6 to Q9, the target
mask is indicated by the red squares, the black cross gives the
expected star location in KIC converted from the star's celestial
coordinates.
}
\label{f:v621lyrTPD}
\end{figure}

One source of the differences in the flux levels stems from the
variable mask sizes used to extract the photometric data for the
SAP light curves \citep{garcia_2011} and the quarterly change of
the CCDs and their different characteristics. Fortunately, the
Kepler archive also includes target pixel data (TPD) that allow one to
check the automatically chosen mask for an individual star. As
mentioned above, the size of the mask is not constant for the four
quarters and is not always covering the target correctly.  For
illustration, we present some example TPD around the star V621 Lyr
(KIC 2570059) in Figure\,\ref{f:v621lyrTPD}.  In the four quarters
Q6 to Q9 we see a shift of the expected center of the star indicated
by the black cross and the actual central pixel. Consequently the
target mask, indicated by red squares, does not include the star's center.
Therefore, neither the SAP nor the PDC light curve show the correct
light variation of this star.  Using TPDs we could correct this
mismatch, and by extracting the data from missing pixels in Q6 and Q9 from
the superstamps segment 930 (see below) we could even complete the
TPD data set.  Furthermore we also extracted and added the missing
quarters Q2 to Q5 and Q10 to Q13 to get a much longer time coverage
for this star.

\begin{figure}[htb!]
\centering%
\includegraphics[width=0.8\textwidth]{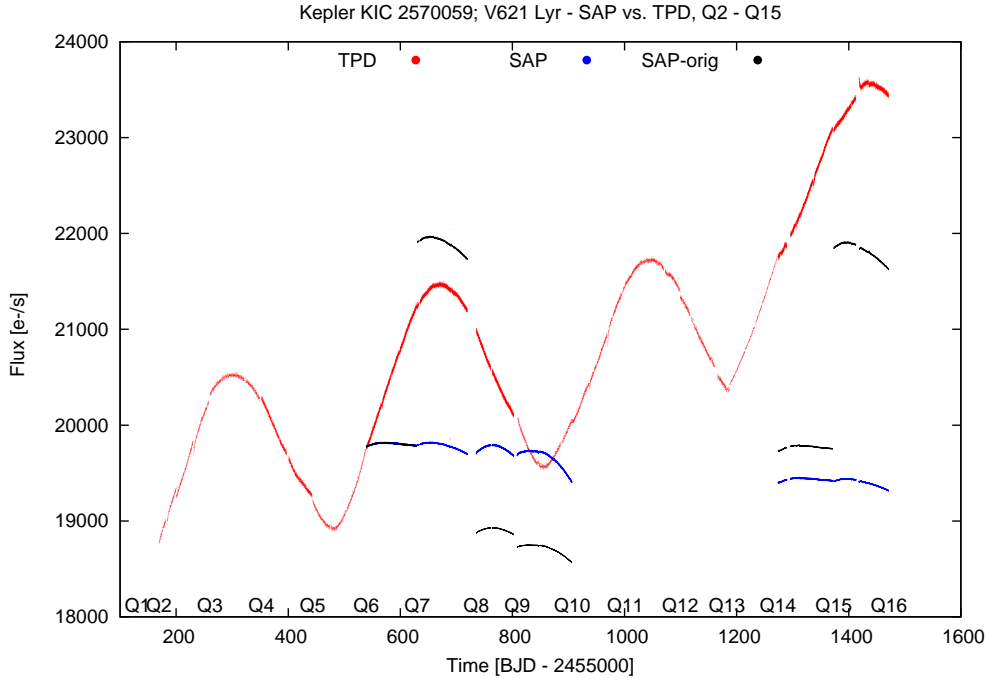}
\caption{Comparison of different data sets for V621 Lyr: SAP (open triangles), SAP zero point corrected and rescaled (crosses), and TPD 
extended with superstamps data from Q2 to Q5 and Q10 to Q13 (filled circles).}
\label{f:v621lyrSAP}
\end{figure}

As a consequence of these mismatches we decided to use a flexible
aperture introducing a threshold value to select the pixels included
for each individual observation. The threshold value was based on
the flux of pixels not belonging to the star, which typically have
significantly lower values and show almost no variation.  Depending
on the star's position in a crowded field or not threshold values between 45
to 220\,e$^{-}$/s give the best results.  In the case of the example
shown in Figure\,\ref{f:v621lyrTPD}, we see that there are four to
eight more pixels significantly contributing to the total light,
compared with the Kepler default target aperture.  A careful
definition of the aperture is therefore of critical importance to
avoid systematic errors in the photometric data.  Our approach
permitted us to reduce strongly the influence of the telescope jitter
on the extracted light curve.  Figure\,\ref{f:v621lyrSAP} compares
the result of our approach using TPD and the SAP data directly from
the archive with and without amplitude rescaling.

Linking of the various quarters unambiguously can also remain a
challenge in the case of the TPD data.  A simple connecting, as
described above, can lead to results that are likely incorrect.
This problem occurs primarily for the gap between Q7 and Q8 which
is much larger than the typical ones lasting for more than 15 days.
Here we use the light curve shape from the nearby cycles shifted
to lie over the gap to estimate the needed adjustment immediately
after the gap.


The various sources producing the offset between quarters are a
mixture of multiplicative and additive contributions that can not
be disentangled.  To estimate the resulting range of uncertainty
we tested both approaches, i.e. a purely multiplicative and a purely
additive correction.  For the period determination, both methods
give about the same results, but the amplitude variations are
naturally different.  For a low flux star like V621 Lyr the amplitude
difference between rescaling using a multiplicative factor or using
an additive constant is about $\pm$2\%, whereas for a highly variable
and high flux object like V1766 Cyg we notice changes up to 3.76\%.
In the following we assume that the major effects are of multiplicative
nature and use this for scaling.

While the TPD achieves a better result in some cases problems can
still occur.  For instance, for Q6 to Q9 the row of pixels adjacent
to the center of V621 Lyr is missing in the TPD data.  In such cases
the superstamps discussed in the next paragraph can provide a
solution. Data for missing TPD observations also can be recovered
from the superstamps.  The superstamps for NGC 6791 consist of 20
images of 100 x 20 pixels in the format of the target pixel data
at the 29.4 minute long cadence (LC) intervals. Several of our
targets in the constellation Lyra have superstamp data available.
We used these data to close gaps in the time coverage of the SAP
and TPD light curves. To make sure we can combine the superstamp
data with the other data sources we extracted quarter 10 data of
V607 Lyr, where we have also TPD data for the same time. The agreement
is excellent.

One disadvantage of this method lies in the TPD like structure of
the superstamp data. For one complete set of 100 x 20 pixels only
one quality flag is available, and in many cases it shows values
far above the good quality limit of 1000. For the TPD analysis we
ignored all observations with non-zero quality flags.  For the
superstamps the same restriction results in just a handful of
observations in a quarter.  An acceptable solution was found by
accepting observations with a summed up quality flag up to 9000 and
adding some additional formal checks.  For instance we remove data
with zeros in the date or flux and extreme outliers identified by
a two-point difference function. A few obviously wrong data still
in the light curve after these checks had to be removed manually.

Our analysis revealed that the best way to construct complete and
reliable light curves from the Kepler data for long period variables
is to apply a mix of methods combining various types of data provided
in the Kepler archive.  In difference to the work of 
\citet{Banyai13} we use TPD data instead of SAP data in our analysis.
 
\subsection{Light curve analysis}

\subsubsection{Fourier methods}

As a primary period search method we used classical Fourier analysis
combined with a least square method optimization provided by the
Period04 software package \citep{lenzbreger2005}. The search was
limited to a maximum frequency of 0.2 cycles per day, i.e. a lower
period limit of 5 days. Up to 4 periods were used to fit the light
change. While this upper limit for the number of periods included
in the model fit was set somewhat arbitrarily, it was well motivated
by our experience that including further periods usually does not
lead to a significant improvement of the fit. Furthermore, in many
of our sample SRVs we observed a clear drop of the S/N ratio for
additional periods in the Fourier power spectrum (see also Table
\ref{t:pstars}). However, the
existence of further significant periods cannot be totally excluded.
The Fourier analysis was done on the light curves without co-trending.

\subsubsection{Autocorrelation}

In addition to Fourier analysis, other methods have proven useful
in light curve analysis, revealing additional information beyond
the periods present in the light curves \citep{tempelton_2004}.
For stars with part of the fundamental group description including
irregularities in the light curve autocorrelation and related
self-correlation methods have been found to be useful in quantifying
the cycle-to-cycle variation in addition to the finding the dominate
period \citep{percy_mohammed_04,percy_et_al_2009}.  A regularly
periodic light curve will have strong correlations for time lags
that are multiples of the period with the strength of the peak
correlation staying high over many cycles. Semi-regular stars will
generally have a strong correlation for the first few multiples of
the period but the coefficient drops off over time.  The persistence
of the autocorrelation can then be used as an indicator of the
regularity present in a semi-regular star. Light curves which have
multiple periods will tend to have weaker autocorrelation peaks
although there can be complex patterns emerging from the beat
phenomena between the dominate periods.

We performed an autocorrelation analysis based on the algorithm of
\citet{autocor}. The traditional autocorrelation routine requires
a perfectly evenly spaced time series, therefore we used an
autocorrelation routine adapted for astronomical time series developed
by Matthew Templeton at the AAVSO (private communication). For the
longer period SR stars, we used a lag timing step of 0.5 days and
maximum lag times up to 1000 days. While lag times larger than half
of the total data range have more limited accuracy, we include these
longer lag times for some stars as an indication of the possible
longer term behavior but always recognizing that the accuracy is
diminished. The autocorrelation coefficient can range from +1.0 for
a perfectly correlated signal to -1.0 for a perfectly anti-correlated
signal \citep{tempelton_2004}.  In this analysis we consider a
strong correlation to be one with a peak coefficient above 0.75
while levels below 0.25 are classified as weak correlation.  In our
analysis of these stars, the periods found using autocorrelation
are generally in agreement with the periods found through Fourier
analysis but we additionally comment of the regularity in the light
curves as measured through the strength and persistence of the
autocorrelation coefficient.


\subsection{Test Sample} \label{s:results}

The resulting periods (numbers of brackets give the S/N ratio derived
from the Fourier analysis) and supplemental data for each star are listed
in Table \ref{t:tstars}. In the text we will refer to the periods
as P1 to P4, listed left to right in the tables, with P1 having the
largest amplitude.  Transformation of Kepler fluxes into Kepler
magnitudes was done using the relation given on the Kepler
webpage\footnote{http://keplerscience.arc.nasa.gov/CalibrationSN.shtml}.

Previous analyses of Kepler data \citep[e.g.][]{Banyai13} identified
a signal of approximately one year length in most light curves.
This so called ``Kepler-year" of 372.5 days length has an origin that has not been fully
explained.  It is obviously related to the orbital motion of the
satellite.  Focus changes due to a varying satellite temperature
are a possible cause\footnote{Kepler Data Release Notes 14:
http://archive.stsci.edu/kepler/release-notes/release\_notes14}. 
This could be augmented by seasonal
variations of the background zodiacal light. The Kepler-year is prominent in
our test sample variables which have low amplitudes.  However, for
the program stars, all of which have large amplitudes, the Kepler-year
has either a minor impact or could not be detected at all. We
performed additional test reductions on stars in Fig.\,5 of
\citet{Banyai13} showing the Kepler year.  Using TPD data instead
of SAP data, the Kepler-year variations disappeared in five of eight
stars and thus could be partly an artifact introduced by the SAP
reduction pipeline.  A detailed investigation of this difference
could shed more light on the origin of this feature in the Kepler
light curves but is beyond the scope of the present paper.

\begin{deluxetable}{llrrrrr}
\tabletypesize{\footnotesize}
\tablecaption{Test Data Results.\label{t:tstars}}
\tablewidth{0pt}
\tablehead{
\colhead{Kepler} & \colhead{GCVS} &  \colhead{GCVS}       &            P1 (S/N\tablenotemark{a}) &           P2 (S/N) &             P3 (S/N) &           P4 (S/N) \\
\colhead{ID}     & \colhead{Name} &  \colhead{Period [d]} & \colhead{[d]} & \colhead{[d]} & \colhead{[d]} &\colhead{[d]} \\
}
\startdata
2437359 & V616 Lyr &  20.9  & 370.11 (8) & 189.10 (1) & long   &  16.5 (1)   \\
2569737 & V607 Lyr &  9.6   & 371.31 (14) & 185.36 (6) & 487.38 (4) & 13.54 (3)   \\
2570059 & V621 Lyr &  10    & 365.31 (7) & 841: (3)   & long   & \nodata  \\
\enddata
\tablenotetext{a}{S/N of the peak in Fourier power spectrum.}
\end{deluxetable}

\subsubsection{V616 Lyr = NGC 6791 V73 = KID 2437359}

\begin{figure*}[htb] 
\centering%
\includegraphics[width=0.8\textwidth]{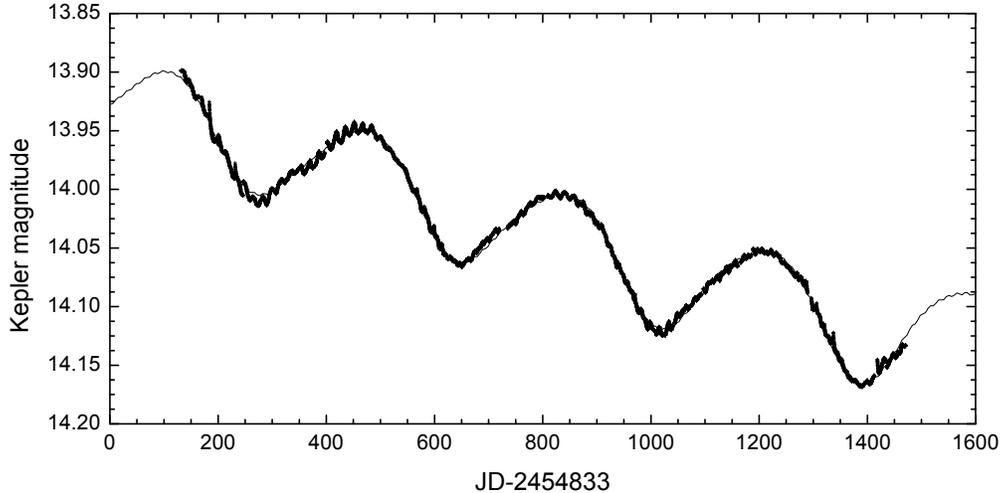}
\caption{Kepler light curve of V616 Lyr. Heavy line data, thin black line is best fit model described in the text.}
\label{V616Lyrplot}
\end{figure*}

A combination of a long trend ($>$5000\,d) and P1$=$370.11\,d,
resulting from Fourier analysis, gives an excellent fit to the light
curve (Fig.\,\ref{V616Lyrplot}).  Adding P2$=$189.1\,d, which is
approximately half the value of P1, gives a visible improvement.
However, P1 is very close to the Kepler year and P1 and P2 are almost certainly
not intrinsic variations of the star.  Cotrending was very
effective in removing this long period.
Also obvious in the data is a very short period of $\sim$16 days.   
\citet{MSK2003}  and \citet{BGT2003} found similar periods. Autocorrelation
and Fourier analysis both show that the 16-16.5\,d period is not a
single period. The typical full amplitude of the 16 d period
variation is around 0.01 magnitudes with the amplitude
changing over time.

\subsubsection{V607 Lyr = NGC 6791 V97 = KID 2569737}

\begin{figure*}[htb!] 
\centering%
\includegraphics[width=0.8\textwidth]{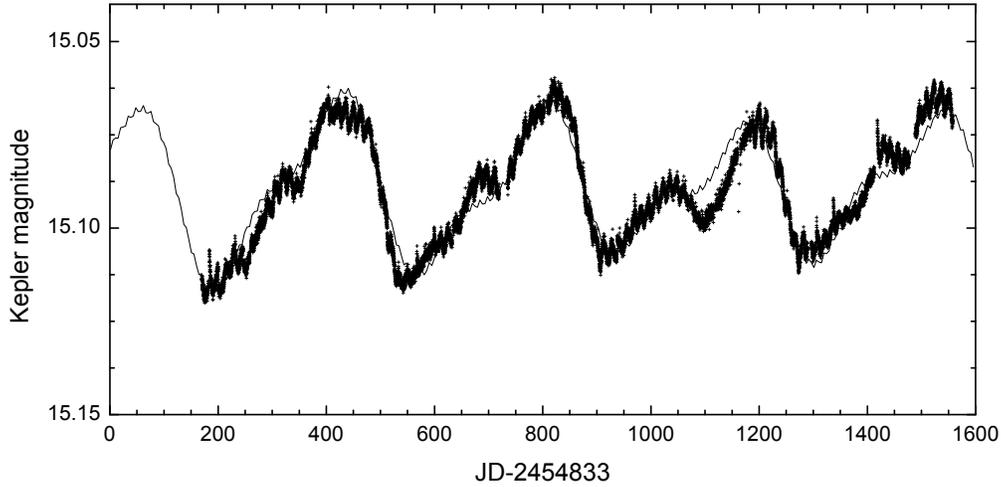}
\caption{Kepler light curve of V607 Lyr. Symbols as in Fig.\,\ref{V616Lyrplot}}
\label{V607Lyrplot}
\end{figure*}

The strongest signal resulting from Fourier analysis is a 372 days
(P1) period.  The fit can be improved by adding essentially the
half value of P1, P2$=$185.36\,d, and P3$=$487.38\,d. Since P1 is
very close to the Kepler year, P1 and P2 are almost certainly not
intrinsic to the star.  The origin of P3 is not clear.  As with
V616 the data contain an obvious very low amplitude variation (P4).
In the case of V607 Lyr the period is $\sim$13.5 days.  P4 agrees
with the 13.6 d period from the ground-based data of \citet{MSS2005}
but differs from the 9.6 day value of \citet{BGT2003}.  The typical
full amplitude of the oscillation is roughly 0.01 mag with the amplitude
changing with time. 
The light
curve shown in Figure \ref{V607Lyrplot} contains a gap over quarters
11 and 12 when it was not observed as a target aperture with Kepler.
We filled that gap using superstamp data as described in
Sect.\,\ref{quarters}. It is interesting to note that the short
period variation is much less obvious in these data. A reason for
this difference is not understood yet. 

\begin{figure*}[htb!] 
\centering%
\includegraphics[width=0.8\textwidth]{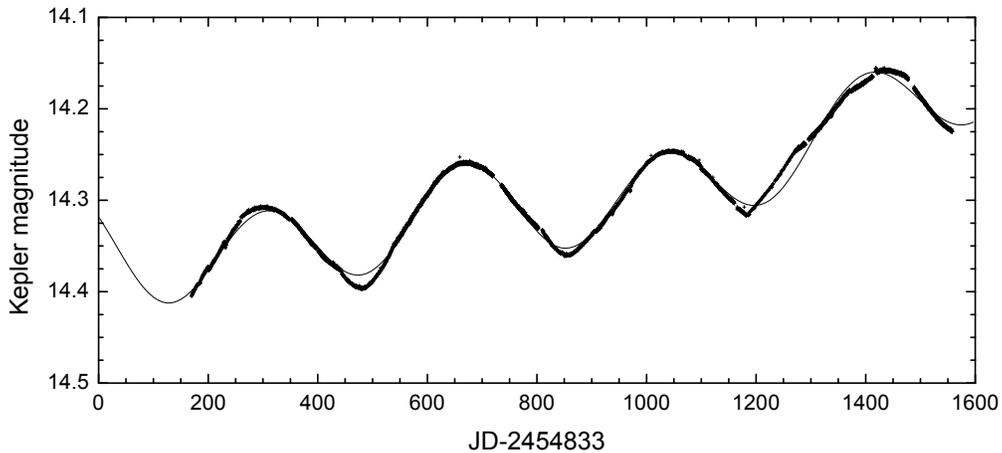}
\caption{Kepler light curve of V621 Lyr. Symbols as in Fig.\,\ref{V616Lyrplot}}
\label{V621Lyrplot}
\end{figure*}

\subsubsection{V621 Lyr = NGC 6791 V99 = KID 2570059}

The amplitude of the Kepler light curve (Fig.\,\ref{V621Lyrplot})
is rather small.  A one year period (365.31\,d, P1) is very clear
but could be identical with the Kepler year.  A combination of P1,
P2$=$841.33\,days, and a long time trend, clearly exceeding the
total length of our time series, gives a very good fit to the Kepler
light curve.  P2 and and the additional long time trend are not
well defined.  To further examine the data we applied cotrending
to the SAP light curves for the individual quarters. A satisfactory
cotrend fit required the first five cotrending basis vectors in
this case since the third and fourth basis vector still seemed to
show some of the original unwanted thermal features. The cotrend
fits matched the shape of the variations of V621 Lyr very well.
Autocorrelation and Fourier analysis of the cotrended light curves
reveal no significant short period signals.  The very small amplitude
$\sim$10 d variations reported by \citet{BGT2003} and \citet{MSS2005}
were not present in agreement with the results published by
\citet{MPM2007}.  The Kepler data for V621 Lyr contain no
intrinsic variations, i.e. at the time of the Kepler mission 
this object was not a variable star.


\subsubsection{Summary of Test Results}

\begin{figure}[htb!] 
\centering%
\includegraphics[width=0.8\textwidth]{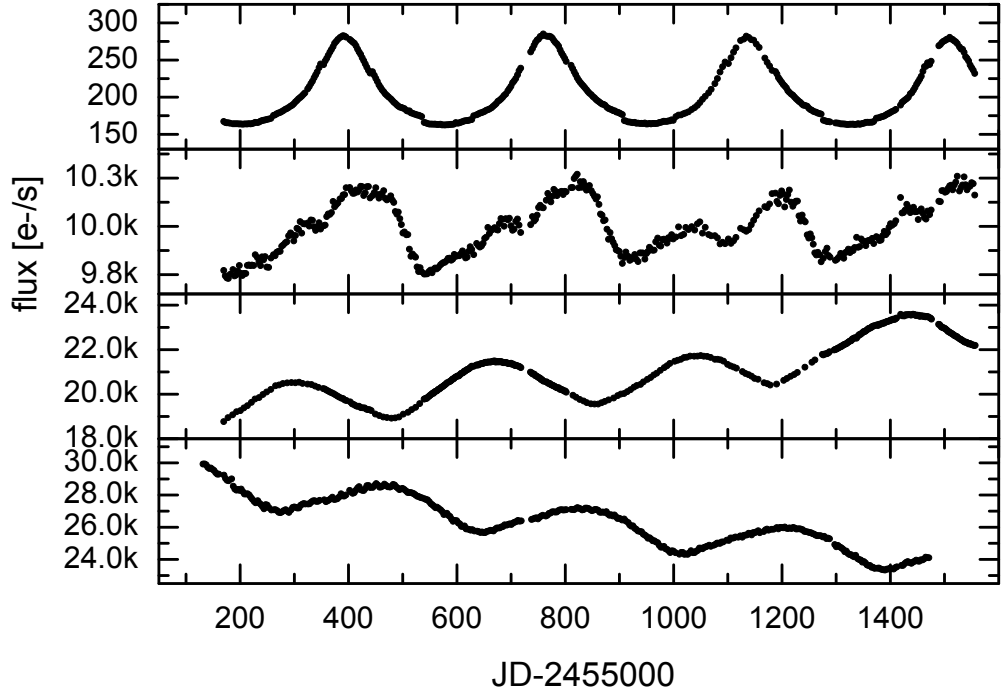}
\caption{Comparison of the Kepler-year for the three
different test stars with low intrinsic variations. The Kepler background
level extracted from V607 Lyr by the Kepler pipeline is also illustrated.  
}
\label{Vxxx_Lyr_kpyrI}
\end{figure}

In the three very low amplitude test stars the Kepler-year artifact
provides the dominant signal.  Figure \ref{Vxxx_Lyr_kpyrI} shows
the TPD light curves for the three low amplitude test stars and the
zoomed and shifted Kepler background\footnote{ A calibrated on
flight sky background is taken from a pixel mask per CCD and the
stars values are calculated by a polynomial expression
\citep{caldwell2010}.  The background consists mainly of two parts:
the seasonal variation of the background flux which is due to the
changing zodiacal light signal and secondly the additionally diffuse
starlight from faint background stars \citep[Figure 6 of][]{caldwell2010}.
} extracted from V607 Lyr.  The Kepler-year periods in these stars
differ by at most 2\% from the nominal Kepler-year period of 372.5
days with amplitudes of less than 0.3 Kepmag.  The Kepler-year
signals are not in phase and all have specific Kepler-year distortions
even when recorded on the same CCD.  This agrees with the findings
on the Kepler-year of \citet{Banyai13}.

In addition to the 372.5 d Kepler year our data are frequently
better fit by including an instrumental period of a half Kepler-year
and periods longer than the Kepler-year.  In addition, a long time
trend at the 0.1 Kepler magnitude scale with a period not defined
in the 4 years of Kepler data is seen in all three stars.  In the
test stars we were able to detected intrinsic periods in the range
13 to 20 days with amplitudes as small as 0.005 Kepler magnitude.


\section{PROGRAM STAR ANALYSIS}

Here the analysis of the individual program stars is discussed.  We
include brief summaries of what we know about these stars from the
literature.  The resulting periods and supplemental data for each
star are listed in Table \ref{t:pstars}.

\begin{deluxetable}{rcccrrrr}
\tabletypesize{\footnotesize}
\tablecaption{Program Star Results.\label{t:pstars}}
\tablewidth{0pt}
\tablehead{
\colhead{Kepler}& \colhead{GCVS} &  \multicolumn{2}{c}{Period} & \colhead{P1 (S/N)\tablenotemark{a}}       & \colhead{P2 (S/N)}  & \colhead{P3 (S/N)}  & \colhead{P4 (S/N)} \\
\colhead{ID}    & \colhead{Name} & \colhead{GCVS [d]}          & \colhead{ASAS [d]} & \colhead{[d]} & \colhead{[d]} & \colhead{[d]} & \colhead{[d]} \\
}
\startdata
3431126  & EG Lyr    & 236                & \nodata & 228.9 (16) & 258.2 (13) & 137.7 (5) & \nodata \\
5176879  & BU Lyr    & 259                & \nodata & 156.5 (32) & 158.9 (31) & 347.5 (6) & \nodata \\
5614021  & V588 Lyr  & \nodata            &\nodata  & 116.4 (9) & 217.9 (9) & 198.3 (6) & \nodata \\
6127083  & V1670 Cyg & 188                & 193     & 187.2 (17) & 94.0 (4)  &\nodata & \nodata \\
7842386  & V1766 Cyg & \nodata            & 113.7   & 122.2 (6) & 754.0 (6) & 104.2 (4) & \nodata \\
8748160  & V1253 Cyg & \nodata            & 187     & 197.2 (13) & 857.4 (3)  & 99.4 (2)  & \nodata \\
9528112  & AF Cyg    & 92.5, 175.8, 941.2 & \nodata & 177.4 (6) & 1867. (8) & 93.6 (5)  &  441.3 (3) \\
10034169 & V2412 Cyg & 250                & \nodata & 255.9 (16) & 241.2 (8) & 140.4 (2) & \nodata \\
12215566 & V1953 Cyg & \nodata            & 95.2    & 163.1 (6) & 847.5 (6) & 93.4 (4)  & \nodata \\
\enddata
\tablenotetext{a}{S/N of the peak in Fourier power spectrum.}
\end{deluxetable}



\subsection{EG Lyr (KID 3431126)}

\citet{PKS63} found that EG Lyr had an M5 III spectrum and rejected
the earlier identification as an RV Tau variable.  \citet{miller_1969}
classified EG Lyr (using the alias VV 221) as a LPV with an uncertain
period of 267 days.  The 4th edition of the GCVS \citep{kholopov_1985}
reclassified EG Lyr as an SRb star with a period of $\sim$236 days.
\citet{wahlgren_1992} found the colors to be in agreement with an M6 III.
The ASAS observations of EG Lyr cover about two cycles with a period
roughly consistent with the GCVS value.  However, ASAS classified
the star as aperiodic.

\begin{figure*}[htb!] 
\centering%
\includegraphics[width=0.8\textwidth]{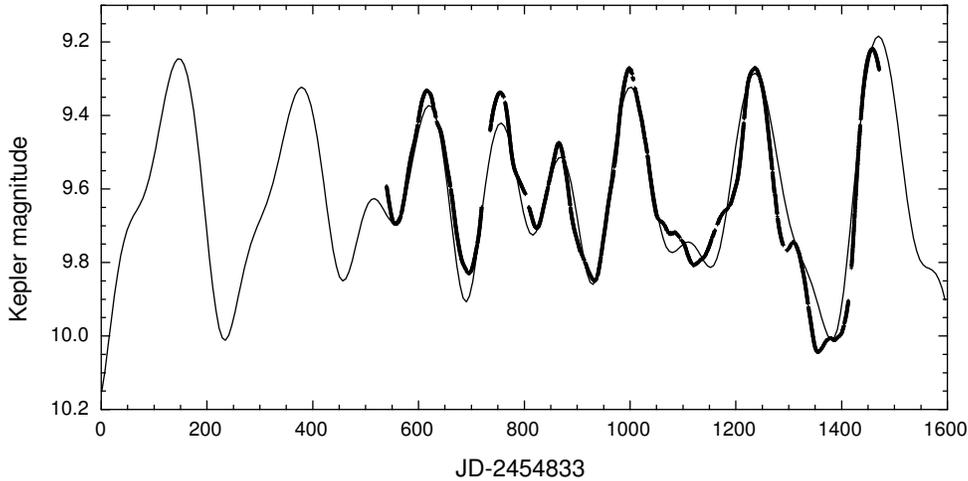}
\caption{Kepler light curve of EG Lyr.  Symbols as in Fig.\,\ref{V616Lyrplot}
Note that the data span a shorter interval than for most of the other targets.}
\label{EGLyrplot}
\end{figure*}

The Kepler data show a strong light variation (see Fig.\,\ref{EGLyrplot}) but
the periodicity is not obvious.  The full amplitude of the
oscillations, from minimum to maximum, range from about 0.2 to 0.8
magnitudes.  In general the peaks in the light curves are fairly
narrow across the whole light curve.  The minima are fairly narrow
in the first half of the light curve but become much wider with
evidence of greater structure in the second half of the light curve.

Three periods, P1=228.9 d, P2=258.24 d, and P3=137.74 d, give a
reasonably good fit and predict the times of the observed maxima
and minima as well as the overall shape of the light curve.  After
subtraction of the periods the residuals suggest the existence of
a long time trend, considerably longer than the observed time series,
and a short period with a length around 78 days but without a strict
periodicity.  Autocorrelation analysis similarly shows a strong correlation
at 238 days with a less well resolved period at 140 days.  The GCVS
period of 236\,d is a good match to the autocorrelation period and
falls between our Fourier periods P1 and P2 providing overall good
agreement on the period length.

\subsection{BU Lyr (KID 5176879)}

BU Lyr was initially identified as a variable by \citet{ross_1928}
from a one magnitude difference in brightness measure 20 years
apart.  It subsequently was identified by \citet{neugebauer_leighton_1969}
as a bright near-IR star, $K=$2.95, with an infrared index $(I-K)=$4.39.
\citet{zverev_makarenko_1979} confirm that this is a very late-type
star with a spectral type of M7 and a stable mean period of 259
days.  The light curve appeared to have a double maximum.  The
brightness of the main maximum is fairly stable but the brightness
and phase of the second maximum and the minima is variable.  However,
ASAS observations of BU Lyr show variability without a clear periodic
pattern and this survey classified BU Lyr an aperiodic variable.
The GCVS classifies BU Lyr as an SRb variable with a visual amplitude
of 1.4 magnitude.

\begin{figure*}[htb!] 
\centering%
\includegraphics[width=0.8\textwidth]{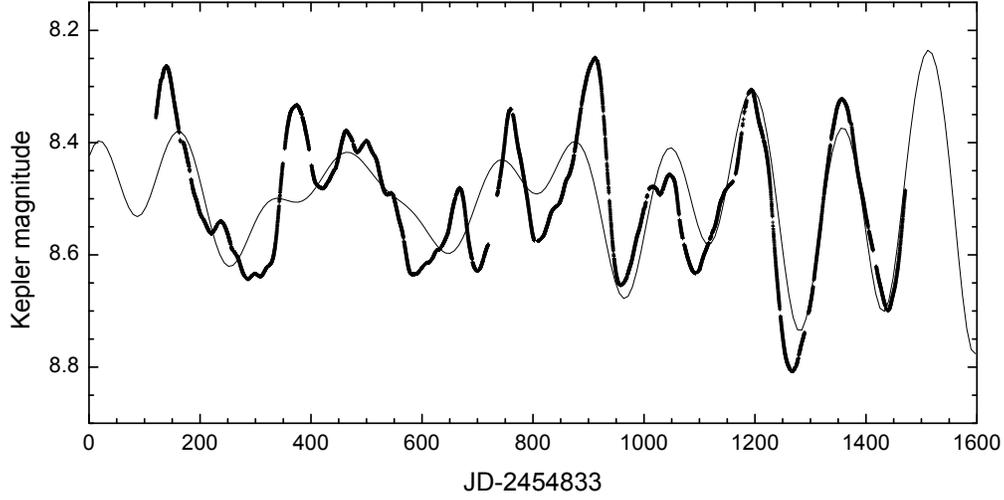}
\caption{Kepler light curve of BU Lyr.  Symbols as in Fig.\,\ref{V616Lyrplot}}
\label{BULyrplot}
\end{figure*}

The Kepler light curve looks irregular (see Fig.\,\ref{BULyrplot})
but the amplitude is quite large.  Typical variations are in the
general range of 0.3 magnitudes with variations just over 0.5
magnitudes for the full light curve.  Autocorrelation analysis
suggests a weakly correlated period of about 150 days.  Fourier analysis of the data
set gives two periods very close together: 156.52\,d (P1) and
158.90\,d (P2).  Adding a variation on a time scale of about one year
(P3) reproduces some aspects of the light curve including the correct
prediction of most of the maximum and minimum times. However, much
of the shape is not properly reproduced. No clear periodicity is
visible in the residuals after subtracting P1 to P3.  The double
maximum structure reported by \citet{zverev_makarenko_1979} is not
seen in the Kepler data. Instead, the brightness of the maxima and
minima seems to change in an irregular way.  Similarly we do not
see a 259\,d period.  The residuals O-C over the time span of our
Kepler observations suggest a tendency towards an increasing period.

\subsection{V588 Lyr (KID 5614021)}

\citet{dahlmark_2000} identified V588 Lyr as a possible Mira of
period 221 days and a range from 12.5 magnitudes to 14.0 magnitudes.
The GCVS lists V588 Lyr as an SRa variable. ASAS observations found it to be
aperiodic with no clear light curve shape at the time of observation.

\begin{figure*}[htb!] 
\centering%
\includegraphics[width=0.8\textwidth]{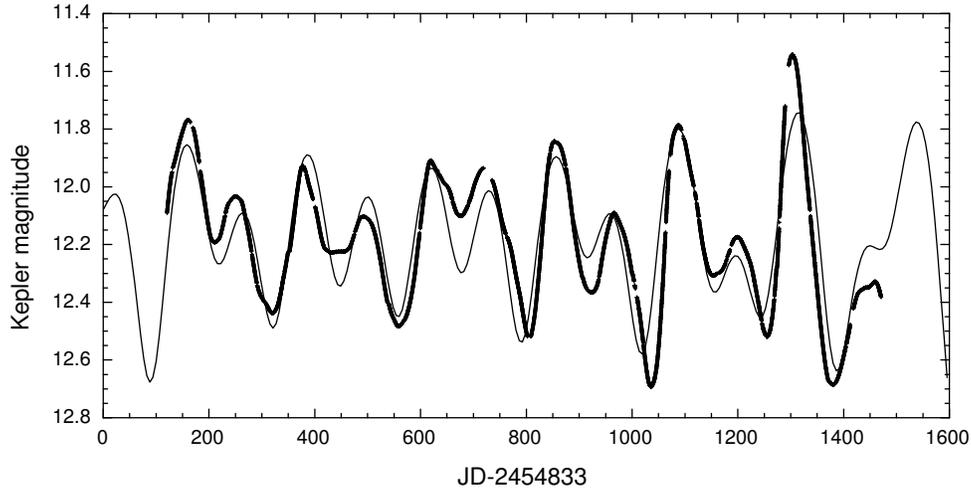}
\caption{Kepler light curve of V588 Lyr.  Symbols as in Fig.\,\ref{V616Lyrplot}}
\label{V588Lyrplot}
\end{figure*}

The Kepler light curve of V588 Lyr (Fig.\,\ref{V588Lyrplot}) shows
strong oscillations covering slightly over 1.1 Kepler magnitude of
variation with individual maxima and minima levels varying considerably
from cycle to cycle. During the time of observation the maxima and
minima had alternate bright and faint cycles.  The combination of
two periods resulting from a Fourier analysis, P1(116.39\,d) and
P2 (217.93\,d), permits a fit to the alternating depths, and the autocorrelation
analysis agrees with these periods with even peaks more strongly correlated
than the odd peaks.  The
addition of P3 (198.33\,d) slightly improves the result but a few
maxima/minima are not well reproduced. The residuals do not suggest
the presence of an additional long period trend or further short
periods, but rather non-sinusoidal variations and additional
cycle-to-cycle irregularities.  P2 is almost identical to the value
given by \citet{dahlmark_2000}.

\subsection{V1670 Cyg (KID 6127083)} 

Hoffleit and Sands \citep{sands_1978} classified V1670 Cyg as an
SR star with a period of 187.5 days.  \citet{dahlmark_2000}
rediscovered V1670 Cyg classifying it as a 191 day Mira with a visual
magnitude ranging from 12.4 -- 15.0.  ASAS found V1670 Cyg to be
quasi-periodic with a period of 193 days and visual amplitude of
1.99 magnitudes.

\begin{figure*}[htb!] 
\centering%
\includegraphics[width=0.8\textwidth]{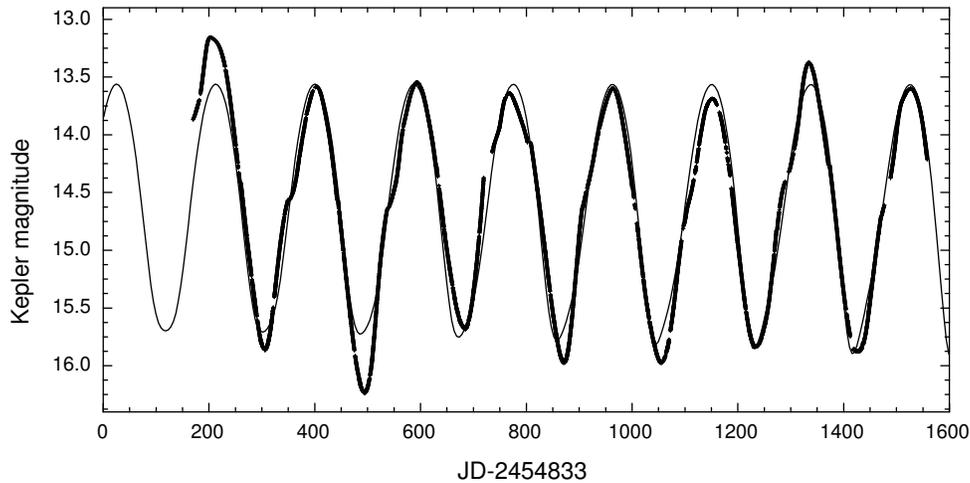}
\caption{Kepler light curve of V1670 Cyg.  Symbols as in Fig.\,\ref{V616Lyrplot}}
\label{V1670Cygplot}
\end{figure*}

The light change observed by Kepler (Fig.\,\ref{V1670Cygplot}) is
very regular except for a very bright maximum at the beginning of
the time series and a second bright maximum shortly before the end
of our time series. There are variations in the depth of the minima
from cycle to cycle.  Autocorrelation indicates a strong correlation at a period of 187.5
days.  Fourier analysis gives a single period of  187.19\,d (P1),
consistent with the autocorrelation analysis. Subtracting P1 from
the light curve reveals a further, rather regular change corresponding
to P2 (94.01\,d) which is very close to half of P1. No further
periodicity was found in the residuals. P1 is in nice agreement
with the literature values for the period.

\subsection{V1766 Cyg (KID 7842386)}

\citet{huruhata_1983} found V1766 Cyg to be a fairly regular SRa
variable with a period of 119 days and a visual light range between
10.3 and 11.5 magnitudes.  V1766 is classified as an MS star
\citep{stephenson_1984}. Based on infrared colors \citet{wang_2002}
suggests that V1766 Cyg is an extrinsic S-star, i.e. its enrichment
in s-process elements results from mass transfer from a more massive
companion. ASAS lists the star as quasi-periodic with a period of
113.7 days and a visual amplitude of 0.80 mags.

\begin{figure*}[htb] 
\centering%
\includegraphics[width=0.8\textwidth]{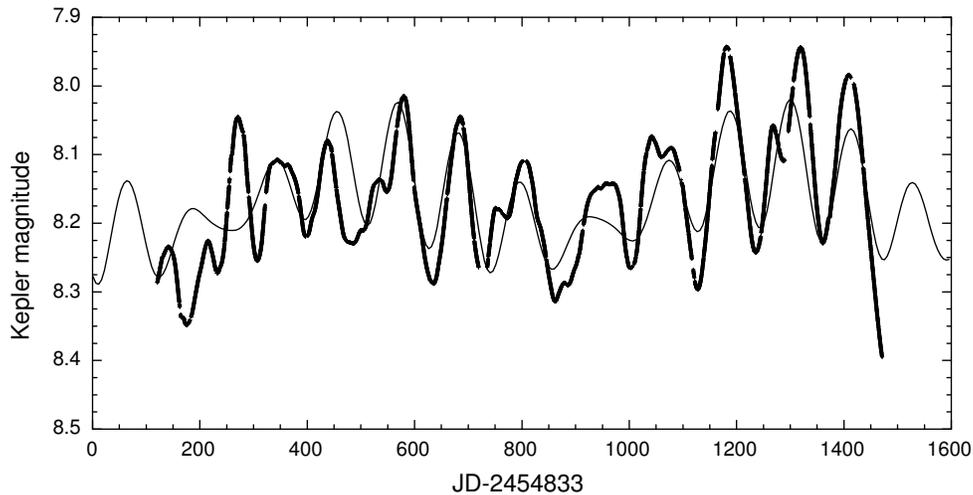}
\caption{Kepler light curve of V1766 Cyg.  Symbols as in Fig.\,\ref{V616Lyrplot}}
\label{V1766Cygplot}
\end{figure*}

In this case, periodicity is not very strongly expressed in the
Kepler light curve (Fig.\,\ref{V1766Cygplot}).  Typical variations
are generally in the range of 0.2 magnitudes with variations just
over 0.4 magnitudes for the full light curve. Autocorrelation
analysis suggests periods of 111\,d and 120\,d with only moderate
to weak correlation.  The Fourier
diagram shows several peaks around 100 days period, a combination
of 122.25\,d (P1) and 104.21\,d (P3) reproduces at least some
features of the light curve. They are close to the ASAS value of
113.7\,d.  Double maxima are present, so we suspect that variations
on a somewhat shorter time scale are present but they do not show
up clearly in the Fourier diagram. A long time trend, which we fit
with a period of 754\,d (P2), seems to be present as well. 

\subsection{V1253 Cyg (KID 8748160)}

\citet{dahlmark_2000} found an uncertain period for V1253 Cyg of
188 to 226 days with a $V$ magnitude range of 11.8 to less than
15th and provided a a classification of SR.  \citet{ASAS} list the
star as quasi-periodic with a period of 187 days.

\begin{figure*}[htb] 
\centering%
\includegraphics[width=0.8\textwidth]{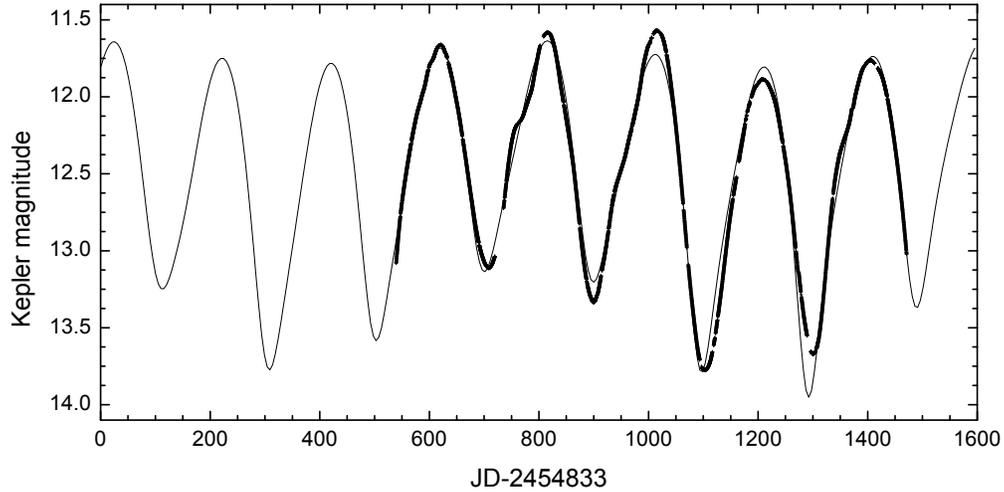}
\caption{Kepler light curve of V1253 Cyg.  Symbols as in Fig.\,\ref{V616Lyrplot}}
\label{V1253Cygplot}
\end{figure*}

The Kepler light curve (Fig.\,\ref{V1253Cygplot}) shows the brightness
of the maxima and minima vary smoothly with time and a bump on the
ascending branch.  The data set is somewhat shorter than the others
covering quarters Q6 through Q15.  Autocorrelation suggests a strong correlation 
with a period
of 196.5\,d.  Using Fourier analysis the times of maxima and
minima can be reproduced very well with a single period of 197.15\,d
(P1). The changing maximum and minimum brightness and the bump
can be qualitatively reproduced by adding a long period of 857.39\,d
(P2) and 99.35\,d (P3).  P2 is, of course, very poorly defined due
to the limited length of the Kepler time series.  P1 is not very
far off the GCVS value. P3 is approximately half the period length
of P1. Its S/N in the Fourier spectrum is only 2.

\subsection{AF Cyg (KID 9528112)}

\citet{fleming_1910} noted that the bright (7.4 -- 9.4 mag) M5 star
now referred to as AF Cyg was variable.  The spectral type at maximum
light was confirmed M 5 III by \citet{keenan_1966}.  The variability
has been extensively studied over the last 100 years.  The principal
period is $\sim$93\,d.  \citet{houk_1963} first commented on the
long secondary period (LSP) of 960 days. Using AAVSO data
\citet{mattei_et_al_1997} found periods of 92.9 and 165.9\,d. An
analysis of four sets of visual observations by \citet{kiss_et_al_1999}
confirmed the periods of 93 and 163\,d and also confirmed a LSP
of 921\,d. The GCVS lists periods of 92.5, 175.8, and 941.2
days.  \citet{andronov_chinarova_2012} discuss the instabilities
observed in the historic light curve. These include alternating
dominance of the 93 and 176\,d periods, intervals of regularly
strong--weak maxima, and variations in times of the 93 day maximum
from 79.4 to 97.4\,d.  For the data analyzed they found a dominant
94.2\,d period with clusters of other periods present.
\citet{hinkle_et_al_2002} observed AF Cyg spectroscopically and
derived an `orbital' period of 926 $\pm$36\,d matching the LSP.
However, they note that the orbits for all LSP systems look the
same so the motion is likely not orbital in origin.
\citet{glass_van_leeuwen_2007} provide a parallax of 4.53$\pm$0.64
mas which allows rough placement on a period luminosity diagram.

\begin{figure*}[htb]
\centering%
\includegraphics[width=0.8\textwidth]{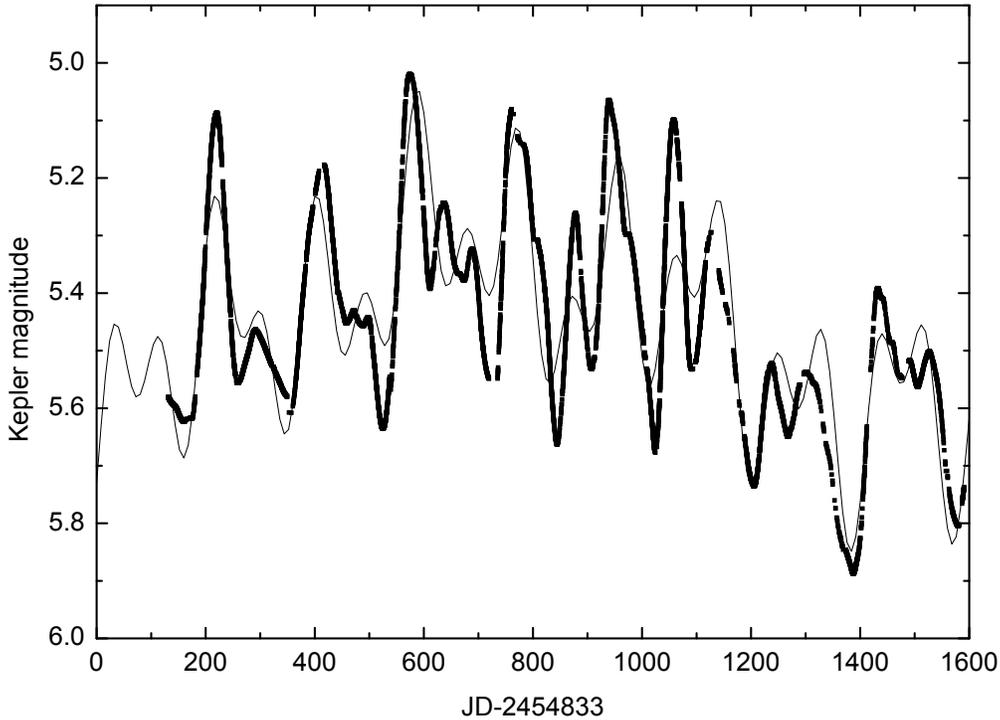}
\caption{Kepler light curve of AF Cyg.  Symbols as in Fig.\,\ref{V616Lyrplot}}
\label{AFCygplot}
\end{figure*}

The Kepler light curve of AF Cyg (Fig.\,\ref{AFCygplot}) shows a
non-strict alternation of strong and weak maxima that could be
interpreted as secondary maxima.  The lack of a strict periodicity
of this phenomenon suggests that instead this is an interplay of
two excited periods of comparable amplitude with a period ratio
close to an integer value.  Several minor bumps are also present
in particular on the descending part of the light curve.  The light
curve appears to have a long time variation exceeding the length of our
monitoring. The typical amplitude of the light change is 0.5\,mag
in the Kepmag system.

A Fourier analysis reveals 4 main periods, and their combination
gives a reasonable fit of the Kepler light curve although
some deviations remain.  The short term variability is accounted
for by P1 (177.4\,d) and P3 (93.6\d). These two periods show a ratio
close to 2, and thus agree with the above mentioned suspicion on
the origin for the alternating bright and weak maxima.  The long
time change is represented by a period of 1867\,d (P2).  While P2
is a strong contributor to the light curve it is not adequately
sampled.  A further significant improvement of the fit can be
achieved by including a fourth period of 441.3\,d.  Similar length
periods were seen in the test data and may be related to the Kepler
year.  Autocorrelation analysis contains significant structure indicating
evidence of multiple periods that are beating against each other. The
highest correlation occurs at roughly 186 days but there are clearly
indications of other periods present. Increasing anti-correlation between
800-1000 days is consistent with the longer period of 1867 days shown in
the Fourier analysis.

For AF Cyg, which has been very well studied from the ground, we
note the agreement of the two well sampled Kepler short periods
with the literature periods (93 and 176\,d) and the disagreement
of the poorly sampled Kepler long periods (441 and 1867\,d) with
the literature period (941\,d).  The 941\,d period is not apparent
in the Kepler data.  A Fourier analysis on the residual light curve
after the removal of P1 through P4 gives an unclear pattern of
additional frequencies between 0 and 0.03 cycles per day (c/d) and
no significant peaks for frequencies above 0.03 c/d.

\subsection{V2412 Cyg (KID 10034169)}

The variability of V2412 Cyg was discovered in 2001 as a part of
the MISAO Project \citep{MISAO} but the only additional information
was a possible coincidence with NSVS 5651487 \citep{nsvs}.  NSVS
5651487 has a light curve with an $\sim$250 day period. ASAS
\citep{ASAS} lists V2412 Cyg as aperiodic.  The GCVS lists a 250
day period for the star and an SR: classification.

\begin{figure*}[htb!] 
\centering%
\includegraphics[width=0.8\textwidth]{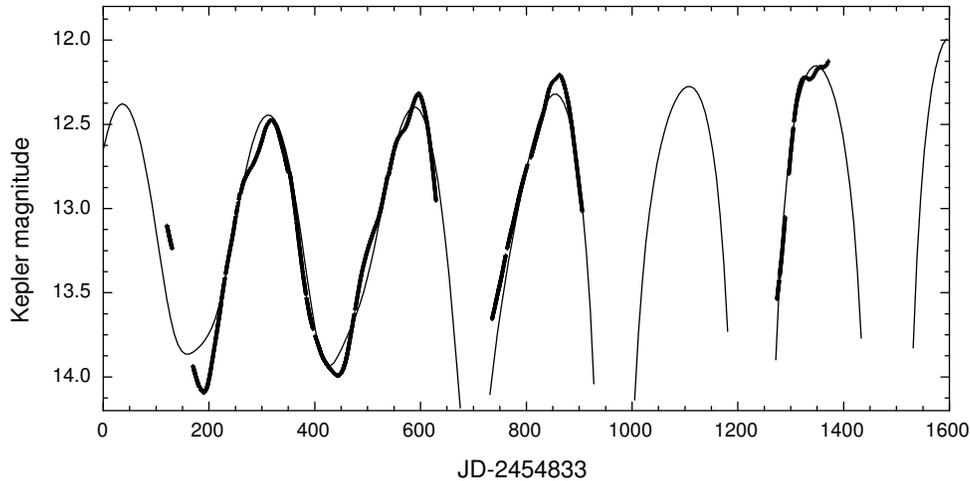}
\caption{Kepler light curve of V2412 Cyg.  Symbols as in Fig.\,\ref{V616Lyrplot}}
\label{V2412Cygplot}
\end{figure*}

The Kepler time series for V2412 Cyg is not continuous (see
Fig.\,\ref{V2412Cygplot}).  This star was observed from quarter Q2
through quarter Q6, then again during quarters Q8 and Q9, and finally
in quarter Q14.  In part the gaps are due to the placement of V2412
Cyg on the failed Kepler Module 3. The large gaps make it extremely
difficult to determine an accurate flux level after each of these
gaps. The behavior during the five continuous quarters gives us
some clues to the proper treatment for this star. During these
quarters we have strong oscillations covering over 1.5 Kepler
magnitudes. The shifts to adjust for discontinuities between these
continuous quarters were relatively small.  We assume that the
corrections in flux for quarters Q8 and Q14 would also be relatively
small and we leave them uncorrected.

Times of maxima and minima can be reproduced reasonably well using
one period of 255.91\,d (P1) only. Variations in the brightness of
the maxima and minima can be fitted nicely using a beat period of
241.19\,d (P2).  A third period of 140.42\,d (P3) comes with a somewhat
low S/N ratio, but including it helps to reduce the residuals
significantly. The light curve shows an asymmetric oscillation with
a steeper descent into minima than the rise to maxima and a slight
bump in the ascending side which seems to shift its position from
cycle to cycle. Including period P3 partly compensates for these
bumps in the rising branches although some difference remains.
Unfortunately an enticing wiggle in the ascending side during quarter
Q14 $(\mbox{JD}-2454833\approx1325)$ is cut off before the star
reaches maximum sometime during the missing quarter Q15.  P1 and
P2 are both close to the GCVS value of 250 days.

\subsection{V1953 Cyg (KID 12215566)}

\citet{gessner_1988} found that V1953 Cyg had slow semiregular
variations with a photographic amplitude of 14.8-15.6 mag and a
period of $\sim$150 days.  The GCVS classifies V1953 Cyg as an SRb.
The ASAS observations of V1953 Cyg list the star as quasi-periodic
with a period of 95.2 days.

\begin{figure*}[htb] 
\centering%
\includegraphics[width=0.8\textwidth]{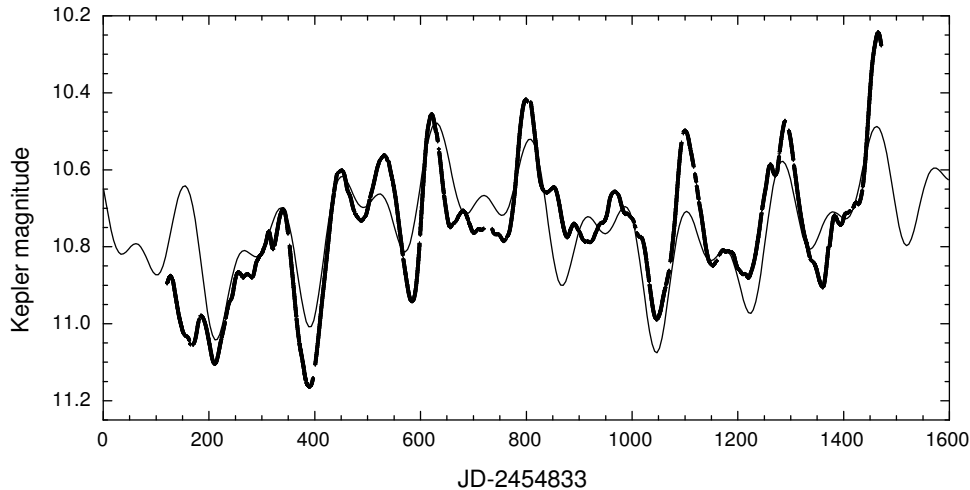}
\caption{Kepler light curve of V1953 Cyg.  Symbols as in Fig.\,\ref{V616Lyrplot}}
\label{V1953Cygplot}
\end{figure*}

Although the Kepler measured light curve for V1953 Cyg
(Fig.\,\ref{V1953Cygplot}) looks quite irregular with a typical
full amplitude of 0.4 mag, a combination of only three periods of
163.07\,d (P1), 847.51\,d (P2), and 93.41\,d (P3) permits a fit to
its main shape.  P2 is poorly defined.  Many details of the light
curve, however, cannot be reproduced properly.  A Fourier analysis
on the residuals gives a number of peaks of similar amplitude between
500 and 50 days.  Autocorrelation analysis finds a moderately correlated period
of 179.5 days, but the correlation decays rapidly indicating that no single period 
dominates the light curve signal.  P3 is similar to the ASAS period.  P1 may be identical
to the 150\,d variation seen by \citet{gessner_1988}.


\section{DISCUSSION}

The Kepler data provide a new way to explore AGB pulsation through
continuous, high-precision, photometric time-series spanning multiple
years without significant interruption.  Diurnal and seasonal effects
in ground-based data result in time series that are too short or
too fragmented for proper handling of the complex variability
patterns observed in stars with very long periods
\citep[e.g.][]{kerschbaum_et_al_2001}.  We were interested in
exploring the various radial modes involved in SR pulsation, looking
for long secondary periods, and searching for the existence of any
low amplitude, short period events or pulsations.  For the latter
goal we were especially interested in looking for any periods that
might be attributable to non-radial pulsations.

The Kepler data, of course, have limitations.  Interruptions of the
Kepler time series observations occurred due to telescope rotation
every three months.  SR variables have periods of order months to
years \citep{GCVS} and extracting periods longer than a Kepler
quarter posed special challenges in data reduction.  Linking the
quarters is surprisingly complicated due to a number of effects
that can introduce photometric errors and we have explored optimum
and non-subjective techniques for the construction of light curves
spanning the whole Kepler monitoring time span.  Another unfortunate
reality of Kepler data, impacting this paper as well as the planet
search goal of the mission, is that most of the stars observed are
moderately faint.  As a consequence most of the variable stars are
not among the best studied objects of their classes.  In spite of
these problems, we find good agreement with the main periods detected
in the Kepler data and ground-based results including the ASAS data
set (see Table \ref{t:pstars}).  The differences appear to be
astrophysical and we will discuss them below.

\subsection{Physical origins of pulsation}

Physical interpretation of the observed pulsational behavior of
red giants has been a topic of research and debate for several
decades. Recent results for low luminosity red giants, based
also on Kepler data \citep{bedding_2010, beck_2011, hekker_2013}, suggest that
solar-like oscillations are observed in these stars.  For the red
giants less luminous than the RGB-tip, three parallel P-L-sequences
are detected \citep{soszynski_et_al_2004} that can be interpreted
as radial low overtone modes plus non-radial (l=1) and (l=2) p-modes
\citep{takayama_2013}.  Asteroseismology using short cadence 
Kepler data provided the possibility of identifying core helium burning
red giants on the basis of frequency separation of gravity modes
\citep{bedding_2011}.

Above the RGB tip where the classical long period variables are found
the pulsation behavior is different.
In the more luminous stars with their larger
light and velocity amplitudes, the role of non-radial modes for the
observed (semi-)periodic behavior is expected to be comparably low
\citep[e.g.][]{wood_et_al_2004}, although some recent results
by \citet{stello_et_al_2014} may initiate a new discussion on this 
question.
Various observational studies, more recently by \citet{Banyai13}
and \citet{kiss_bedding_2004}, have demonstrated the difference in
the variability behavior between RGB stars and the more luminous
AGB stars. As pointed out by \citet{xiong_deng_2007}, stability
analysis of pulsation modes in red giants shows a change from higher
overtone radial modes in low luminosity red variables to lower
overtone radial modes in the more luminous ones.  This is in agreement
with observations from globular cluster giants \citep{lebzelter_wood_2005}.
Model calculations, in particular of Wood and collaborators
\citep[][and references therein]{wood_olivier_2014}, find the
variability of the more luminous stars dominated by radial modes
that are self-excited rather than stochastically excited.  Coupling
between convection and oscillations has been identified as the most
likely excitation mechanism \citep{xiong_deng_2007}.
\citet{buchler_et_al_2004} find that irregular pulsations are driven
by energy exchange between two nonadiabatic modes in a 2:1 resonance.

Pulsation models nicely reproduce the period-luminosity sequences
of luminous red giants \citep[e.g.][]{soszynski_et_al_2007} in terms
of a sequence of the fundamental radial mode and several low order
overtone modes. The only exception is sequence D, the long secondary
periods, which we will discuss in more detail below.  The observed
amplitudes in our program stars place them in the group of luminous
red variables.  The periods and parallax of AF Cyg demonstrate this
unambiguously \citep{glass_van_leeuwen_2007}.  Hence we will focus
on radial pulsation modes in this paper.  \citet{wood_olivier_2014}
find that strange modes may develop in red giants beside the normal
radial modes, but that they are always damped and thus should not
be visible beside the normal radial modes on the upper giant branch.
However, they might affect the period shown by a radial mode.

\subsection{Decomposition of the Light Curve}

The Kepler light curves typically appear at first examination very
complex.  However, in all our stars the combination of two or three
periods was sufficient to predict the times of maxima and minima
correctly.  The exception is BU Lyr where this is correct for only
the second half of the Kepler data set. However, in most cases a
single period was not sufficient.  This implies, as others have
concluded, that even behavior classified irregular is multiperiodic
\citep{lebzelter_obbrugger_2009}.

For the detailed fitting of the complex light change shapes we used
a combination of several periods. The continuity and the high
photometric precision lowered the noise level in the Fourier power
spectra significantly compared with ground-based surveys. In some
cases, the observed changes in amplitudes and period lengths could
be traced reasonably well, while in other stars the large residuals
remained after subtraction of three to four frequencies, just like
in the case of stars well observed from the ground.  This is in
agreement with other studies that note an irregular component in the
light curves \citep{turner_et_al_2010}.

The two independent period search methods we applied lead to very
similar results for period fits to the light curve.  The period
detection is robust for up to four periods at the precision and
time span of the Kepler data.  There are differences between the
periods detected in our study and values in the literature based
on different data sets for the same star.  To explore this further,
we downloaded the ASAS data set for one of our stars, V1953 Cyg,
from the ASAS website\footnote{http://www.astrouw.edu.pl/asas/,
January 2014.}.

We selected V1953 Cyg because its light curve is one of the most
irregular ones in our sample. The fit produced using our Kepler
derived periods can be used to model almost all maxima and
minima in the ASAS data set correctly and the combination of three
periods reproduces a remarkable fraction of the changes in amplitude
and period length. Naturally, the photometric uncertainties of the
ASAS measurements are much larger than those of the Kepler measurements.
Therefore, it is not possible to test the quality of the fit of the
artificial light curve with similar precision.

In the ASAS data base V1953 Cyg is listed with a period of 95.2\,d
although the phased light curve using this period shows considerable
scatter.  Our Fourier analysis of the same ASAS data gives a period
of 95.1\,d as the first peak. An analysis of the residuals results
in a second period of 66.1\,d.  The third period is a long period
with almost 600\,d length.  The Kepler data show a third peak at
93.4\,d and a second peak at 847.5\,d.  There is no Fourier peak
visible in the ASAS data at the prime 163.1\,d Kepler peak.  Attempting
to fit the Kepler light curve using the three periods derived from
the ASAS light curve does not result in a good fit.

We find that the periods derived for semiregular variables are always
connected to the star's behavior at a specific time interval but
are not necessarily transferable to a different data set observed
at a different time.  To identify the dominant period(s) of an SR variable
an analysis of several pieces of the light curve obtained in different
years seems to be necessary.

\subsection{Pulsation Mode and Multiple Periods in the Same Mode}

Multiple periods in SR variables are known from ground-based data.
The Kepler light curves of the SR variables provide extremely clear
examples.  The multiple periods are manifested in three ways.  Most
of the program stars have two pulsation modes present that are
separated by roughly a factor of two (Table \ref{t:pstars}).
Historical light curves show that the dominant mode can switch
between the two excited modes \citep{buchler_et_al_2004}.  In 30
-- 50\,\% of stars a third long period that is $\sim$9$\pm$4 times
longer than the dominant period can also be present
\citep{wood_et_al_1999}.  In addition, each of the two pulsation
modes can be composed of multiple periods spanning a small range
of values \citep{buchler_et_al_2004, soszynski_et_al_2004}.  The
result is that for any given epoch the resulting light curve can
appear irregular or quite periodic.

An excellent example is the SRa variable V1766 Cyg which is described
in the literature as fairly regular with a period of 119 days
\citep{huruhata_1983}.  This is the one star star in our program
that does not have a second mode present.  Nonetheless during the
time of the Kepler observations the light curve appears to be fairly
irregular.  This is due to the presence of a number of periods
spanning the range from 104 to 122\,d.  On the other hand, two stars
in our sample had SR: or SRb classifications, V1253 Cyg and V2412
Cyg, which imply that historically the light curves have been
irregular (or badly sampled).  These stars do indeed have multiple
modes/periods but during the interval of the Kepler observations
light curves were quite regular.

Time variations in the periods of late-type stars pulsating in
multiple modes are well known from comparisons of observed and
average periods.  Changes in the strengths of radial modes also are
a feature of type II Cepheids and RV Tauri stars
\citep{pollard_et_al_2000}.  In these stars as well as in LPVs
variations of this type have been attributed to chaotic processes
\citep{buchler_et_al_2004}.  Using combined sinusoid fits to the
LPV light curves from space missions like Kepler or COROT with a
limited number of periods will not result in vastly different insight
over ground-based analysis of these objects but does produce more
reliable secondary periods.  A substantial collection of high
precision data could form the basis for testing statistical models
of interior processes in LPVs \citep[see for
example][]{buchler_et_al_2004}.

\begin{figure}[htb!]
\centering%
\includegraphics[width=0.8\textwidth]{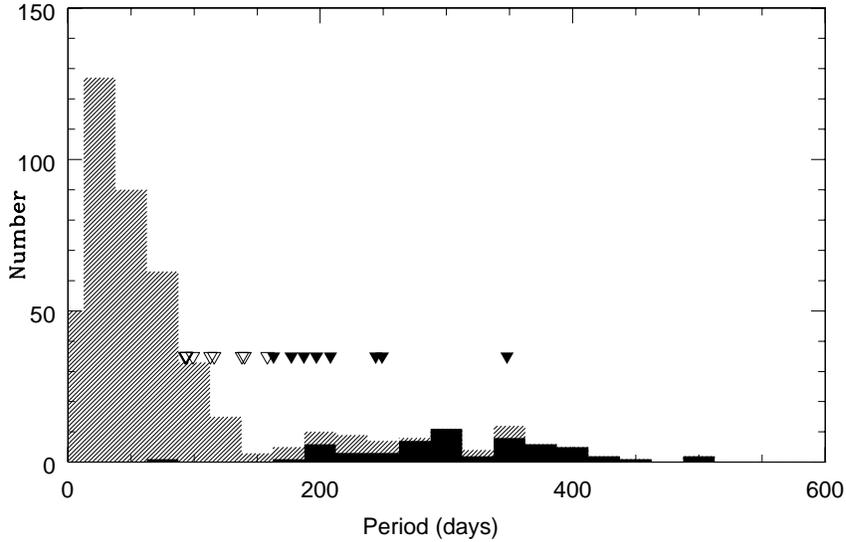}
\caption{
Filled distribution: Kepler field ASAS Miras. Shaded distribution: ASAS Kepler field
QPER variables + Miras.  The filled triangles mark the long periods for the program 
SR stars and the open triangles mark the short period. For V1766 Cyg the $\sim$114 day period
is shown as an open triangle.  Long secondary periods (over 500 days) are not shown.}
\label{f:histogram}
\end{figure}

The ASAS catalog \citep{ASAS} of variable stars in the Kepler field 
lists a total of 947 variables.  Of these approximately 43\% are
in the ASAS category QPER.  The QPER category consists mainly of
SR variables, many with short periods, although we find that some
of our program stars are classified as APER.  Miras make up another
6 percent of the ASAS catalog variables.  Figure\,\ref{f:histogram}
shows the period distribution of the QPER and Mira classes.  A
period distribution is the projection of the period-luminosity
diagram onto the period axis.  We are unable to produce a
period-luminosity diagram since the luminosities of the Kepler field
stars are unknown.  For most of the program stars the
dominant period is around 200 days with a shorter period of around
100 days.  The longer periods match the long period SR distribution
which is shared by the Miras.  The Miras are fundamental mode
pulsators \citep{soszynski_et_al_2013}.  The figure suggests that
for most, if not all, of the program stars the long period is a
fundamental mode.

\begin{figure}[htb]
\centering%
\includegraphics[width=0.8\textwidth]{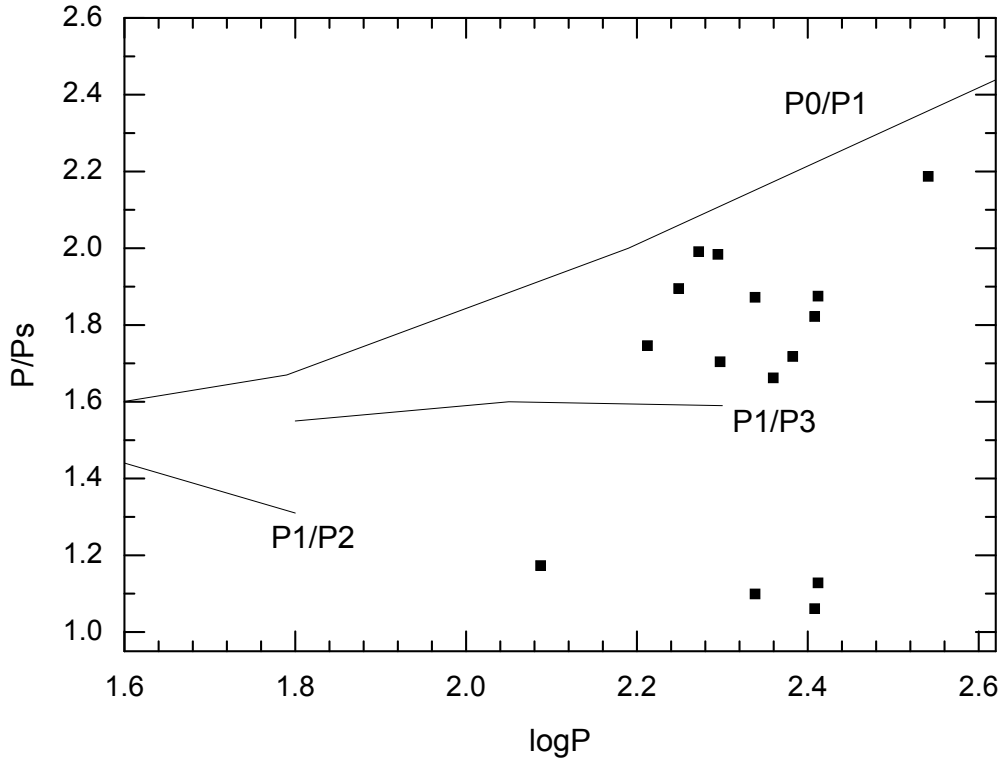}
\caption{Petersen diagram for the periods of the SRVs (Table \ref{t:pstars}).
Period ratios above three (long secondary periods) have been excluded. Lines indicate
theoretical period ratios from \citep{wood_et_al_1999}. }\label{petersen}
\end{figure}

To explore the periodicity patterns detected in our semiregular
variables the period ratios are shown on a Petersen diagram in Figure
\ref{petersen}.  The Petersen diagram sets the ratio of the longer
to the shorter period in relation to the longer period
\citep{Petersen_1973}. This approach has become widely used
for the study of large sets of LPV light curve data 
\citep[e.g.][]{wood_et_al_1999, soszynski_et_al_2013}. In our Figure
\ref{petersen} we see a clustering of data points at period ratios
of 1.1 and around 1.8. The ratios with the very long periods have
been excluded. We show the theoretical ratios from fundamental
mode (P0) and low overtone mode pulsation models (Pn) given in
Figure 4 of \citet{wood_et_al_1999}.  The period ratios around 1.8
fall between the theoretical sequences for fundamental to first
overtone mode and first to third overtone mode, respectively.  All
our sample SRVs with the exception of V1766 Cyg are found in that
region with a pair of periods.  Our result is consistent with the
Petersen diagram presented in \citep{wood_et_al_1999} for long
period variables in the LMC. In analogy to their result we conclude
that most of the SRVs in our sample are fundamental mode pulsators
which also have the first overtone mode excited.

The period ratios clustering around 1.1, i.e. pairs of periods very
close together, are not the result of two different low order radial
pulsation modes \citep[cf.][]{soszynski_et_al_2013}.
\citet{soszynski_et_al_2004} detected that about 35\,\% of the LMC
semiregular variables in the OGLE database show such a behavior.
In analogy to similar observations found in RR Lyr stars and Cepheids,
these authors attribute such periods to non-radial oscillations.
However, non-radial oscillations of the size required for M giant
SR variability would result in unrealistic distortions \citep{wood_et_al_2004,
nicholls_et_al_2009}. Non-radial modes also are not expected to
provide significant velocity variations \citet{wood_et_al_2004}.
In the case of AF Cyg the velocity amplitude is $\sim$ 5 km\,s$^{-1}$.
Similar results are found for the related stellar sample presented
by \citet{lebzelter_hinkle_2002}.

\citet{bedding_et_all_2005}, while excluding non-radial oscillations
as the origin of multiperiodic behavior, pointed out that the
clustering of peaks in the Fourier power spectra of some LPVs has
a striking similarity to the power spectra seen for stochastically
excited pulsators like our sun. These are peaks from the same
pulsation mode but with slightly different periods.  However,
\citet{buchler_et_al_2004} point out that the low mass and high
luminosity of AGB stars rules out a stochastic origin and favored
chaotic pulsation dynamics arising from nonlinear interaction between
two resonant pulsation modes.  The light curves show that these
peaks can co-exist or come and go without changing the overall
brightness.  \citet{fox_wood_1982} noted a possible connection to
convection.  This has been discussed by \citet{buchler_et_al_2004},
\citet{xiong_deng_2007} and most recently by \citet{soszynski_wood_2013}.
\citet{soszynski_wood_2013} note that convection carries most of
the energy through most of the envelope but is poorly modeled
resulting in large computational uncertainties.

\subsection{Long Secondary Period}

As noted in the introduction the SR variables exhibit long secondary
period (LSP) variations which have periods longer than the fundamental
and typically $\sim$8 -- 10 times longer than the dominant pulsation
period \citep{nicholls_et_al_2009}.  While these long periods have
been known for a long time \citep[][and references therein]{houk_1963},
LSP became an astrophysical problem when \citet{wood_et_al_1999}
identified period-luminosity sequences covering fundamental and
overtone radial modes, as well as an LSP sequence, in MACHO
observations of LMC AGB stars.  Stars with LSP are multimode pulsators
with a low overtone pulsation.  In the AGB stars these are SR
variables.

Recent surveys of luminous red giants by \citet{soszynski_et_al_2007}
and \citet{fraser_et_al_2008} show that LSP is common.
\citet{soszynski_et_al_2007} suggests very low amplitude LSPs exist
in up to 50\% of variable AGB stars.  The Kepler mission proved too
brief to reliably measure LSP for any star.  However, in agreement
with \citet{soszynski_et_al_2007} about half of our light curve
fits require a long period of a few hundred days length.  In the
case of V1953 Cyg an LSP of $\sim$600\,d is detectable from
ground-based data.  We tentatively identify LSPs of 800+ days for
this star and 700+ days and 800+ days for V1766 Cyg and V1253 Cyg,
respectively.  These stars appear indistinguishable from the other
SRVs in terms of global parameters like color.  Among the SRVs with
a long period in our sample, only one (V1766 Cyg) shows in addition
a pair of periods of very similar length (104 and 122 days).  The
other four stars with such pairs of periods do not show an LSP,
however, three of them show a third period about half the length
of the two similar periods. The third period of BU Lyr is approximately
two times larger than its pair of periods.

AF Cyg is well known from ground based observations to have a LSP
\citep{houk_1963}.  In our data it shows an interesting characteristic
of LSP not previously reported.  The ground based LSP is $\sim$960\,d
which is also seen spectroscopically \citep{hinkle_et_al_2002}.
However we can not fit, even by attempting to force a fit by hand,
this period to the Kepler data.  On the other hand a period of
roughly half the LSP is present. In the few stars where the LSP has
been monitored spectroscopically the LSP is never the dominant 
photometric signal but is the dominant spectroscopic signal 
\citep{hinkle_et_al_2002,hinkle_et_al_2009}.
Archival photometry of these stars suggests that the LSP can 
at times be difficult to observe.

The physical cause of LSPs remains unknown \citep{nicholls_et_al_2009}.
No clear non-radial modes were detected in our data in agreement
with the previous discussion.  Other investigators have excluded a
number of other causes with remaining explanations including star
spots \citep{wood_et_al_2004} and long-period convective cycles
\citep{stothers_2010}.  Long-period convective cycles were mentioned
as a possible cause discussed above for multiple periods in overtone
modes.  The requirement of dual mode pulsation for a long period
suggests that the mode switching mechanism is involved with the
long period.  Coupling between convection and oscillation in SR
pulsation has been invoked by both \citet{buchler_et_al_2004} and
\citet{xiong_deng_2007}

The large amplitude LPV variables are Miras and these have been
proven to be fundamental single-mode variables
\citep{soszynski_et_al_2013}.  \citet{soszynski_et_al_2013} also
find that most SRVs are double mode variables with the fundamental
mode and first overtone simultaneously excited.  Our data similarly
show most SRVs pulsating in two modes.  As discussed in
\citet{soszynski_wood_2013} AGB stars evolve through instability
stages from overtone mode pulsation to fundamental mode pulsation
as the luminosity increases.  The Miras in the LMC sample investigated
by \citet{wood_et_al_1999}, \citet{soszynski_et_al_2013} and others
are single-mode variables and do not have an LSP.  Since the SR class
contains fundamental mode pulsators, the implication is that either
LSP behavior limits the amplitude of pulsation or that a large pulsation
amplitude somehow excludes LSP.  Both explanations require that LSPs
are an intrinsic feature of an AGB star related to overtone pulsation.
This agrees with the conclusion reached by \citet{wood_nicholls_2009}
using other lines of evidence.


\subsection{Very short period variations in LPVs}

The 30 minute sample interval of the Kepler observations permits a
search for rapid outbursts or very short time variations that have
been discussed for LPVs on the basis of Hipparcos data
\citep{delaverny_1998}.  A detailed search by eye of the program
star Kepler light curves for such outbursts lasting between hours
to a few days did not result in any detections. The time around the
transition between two quarters was excluded. This result adds to
other recent studies of this phenomenon that also did not detect
such variations \citep{lebzelter_2011,wozniak_2004}.  We support
the conclusion of these works that outbursts of duration less than
a day in long period variables are very rare events.


\section{CONCLUSIONS}

Kepler data requires considerable care in reduction when used for
multi-year high precision photometry.  The technical problems have
been discussed at length and solutions are presented.  The resulting
light curves of the SR variables are remarkably featureless at high
time resolution and high precision.  No variations were detected
on `rapid' time scales of one day or shorter.  We have undertaken
various types of period analysis on the data.  While many of the
light curves appear irregular the analysis shows this to be the
result of combining two or three regular periods, suggesting that
even the most irregular late-type variables are multiperiodic SR
variables.

Multiperiodicity is a ubiquitous property of the SR variables that
we analyzed.  Using a Petersen diagram the SR variables in the
current sample can be shown to be fundamental plus first overtone
multimode radial pulsators.  The multiple periods of the program
stars are not solely the result of pulsation in more than one mode.
These stars are in fact multiperiodic in three ways: They are
generally pulsating in both the fundamental and an overtone mode.
They have multiple periods in the dominant pulsation mode. Approximately
half of the stars also have a long secondary period 8 to 10 times
the length of the dominant pulsation period.  The strong connection
between LSP and multimode overtone pulsation suggests that LSP is
related to the convective processes that drive mode switching and
multiperiod modes in SR variables.

One of our goals was to use Kepler data to compare the different
subclasses of SR variables.  All the sample stars are multiperiodic
with no apparent differences between the SRa, SRb, or SR: subclasses.
We interpret the apparent regularity of some light curves as 
the result of constructive interference between beating periods.
The implication, which is not surprising, is that the SR classification
scheme is very insensitive to the location of a variable in the P-L 
diagram.  When Gaia data becomes available it will be possible to
place the Milky Way LPVs on the P-L diagram.  Mira variables have
been separated from the SR variable by pulsation amplitude but are
also single mode pulsators.  Previously SR variables have been seen 
as simply low amplitude versions of the fundamental mode mira
pulsators \citep{hinkle_et_al_1997} but this is clearly an incomplete
description.  It will be of interest to see how the extensively
studied late-type variables fit on a P-L diagram where multimode
versus single mode and large amplitude versus small amplitude 
variables can be compared. 


\acknowledgments
The work of TL has been supported by the Austrian Science Fund under
project number P23737-N16.  EH thanks Mark Taylor for STILTS support.
We wish to thank Steve Howell for his helpful comments and for
providing the Mayall 4 m spectra.  We also thank the Kitt Peak
directors for their allocations of time on the Coude Feed telescope.
Finally, we acknowledge the undergraduate students at SCSU who
participated in the initial analysis of the Kepler data.  Support
to SCSU was provided through the NSF PAARE award AST-0750814 and
NASA Kepler awards NNX11AB82G and NNX13AC24G.  The National Optical
Astronomy Observatory is operated by the Association of Universities
for Research in Astronomy (AURA) under cooperative agreement with
the National Science Foundation.



\end{document}